\documentclass[journal]{IEEEtran}
\usepackage{amsmath,amsthm,amssymb,paralist,subfigure,cite,graphicx}
\usepackage{algorithm2e}
\newtheorem{lem}{Lemma}
\newtheorem{theorem}{Theorem}
\newtheorem{defn}{Definition}
\newtheorem{rem}{Remark}

\def\mb{\mathbf}
\def\mbb{\mathbb}
\def\mc{\mathcal}

\begin{document}
\title{\huge Distributed Estimation Recovery under Sensor Failure}
\author{Mohammadreza Doostmohammadian$^\dagger$$^\S$, \textit{Member, IEEE}, Hamid R. Rabiee$^\dagger$, \textit{Senior Member, IEEE}, Houman Zarrabi$^\ddagger$, \textit{Member, IEEE}, Usman A. Khan$^\ast$, \textit{Senior Member, IEEE}

\thanks{
$^\dagger$ ICT Innovation Center for Advanced Information and Communication Technology, Sharif University of Technology, Tehran, Iran {\texttt{m.doostmohammadian@ictic.sharif.edu},
\texttt{rabiee@sharif.edu}}.

$^\S$ Mechanical Engineering Department, Semnan University, Semnan, Iran.

$^\ddagger$ Iran Telecommunication Research Center (ITRC), Tehran,
Iran \texttt{h.zarrabi@itrc.ac.ir}.

$^\ast$ Electrical and Computer Engineering Department, Tufts University, Medford, USA \texttt{khan@ece.tufts.edu}.}}
\maketitle

\begin{abstract} Single time-scale distributed estimation of dynamic systems via a network of sensors/estimators is addressed in this letter. In single time-scale distributed estimation, the two fusion steps, consensus and measurement exchange, are implemented only once, in contrast to, e.g., a large number of consensus iterations at every step of the system dynamics. 
We particularly discuss the problem of failure in the sensor/estimator network and how to recover for distributed estimation by adding new sensor measurements from equivalent states. We separately discuss the recovery for two types of sensors, namely $\alpha$ and $\beta$-sensors. We propose polynomial order algorithms to find equivalent state nodes in graph representation of system to recover for distributed observability. The polynomial order solution is particularly significant for large-scale systems.

\end{abstract}

\section{Introduction} \label{sec_intro}
Information and signal fusion in multi-sensor systems has recently gained a significant attention in signal-processing literature \cite{sayed11,sayedtu12,kar2013consensus,jstsp,jstsp14,das2016consensus,das2015distributed,asilomar14,Msechu:08,schizas2008consensus,camsap11,icassp13}. Emergence of Internet of Things (IoT) 
and the so called Cyber-Physical Systems (CPS) has motivated 
practical applications, where sensors, and generally smart devices are integrated into the physical systems \cite{sensorsummit}.
Recently a growing interest is directed towards distributed estimation in CPS \cite{das2016consensus,das2015distributed,kar2013consensus,jstsp,jstsp14, sauter:09,nuno-suff.ness,battistelli_cdc,sayedtu12,sayed11}, where the state of the dynamical system is globally tracked by  sensors. The sensors communicate and share their estimations/measurements over the communication network at the same time-scale as the time-evolution of system dynamics. This scenario is known as \textit{single time-scale estimation}, and is privileged with low communication demand and no constraint on local observability of sensors (see \cite{das2016consensus,das2015distributed,kar2013consensus,jstsp,jstsp14} for details), in contrast to kalman consensus filters \cite{olfati:05,Msechu:08}.
In this scenario, the system structure dictates the sensor-network structure.  %\cite{das2016consensus,das2015distributed,sayedtu12,sayed11,kar2013consensus,jstsp,jstsp14, sauter:09,nuno-suff.ness,battistelli_cdc}. 
If the system is structurally full rank, a strongly-connected sensor network is sufficient for stable estimation \cite{jstsp,jstsp14,sayedtu12,sayed11,nuno-suff.ness}. However, more than strong connectivity is required for tracking rank-deficient systems, where it is shown that centralized communication hubs are required to recover observability by the monitoring cyber-network \cite{jstsp,jstsp14,sauter:09}. Indeed, (distributed) observability of the system determines the underlying communication topology among sensors/estimators\footnote{In this paper sensor and estimator are used interchangeably, referring to distributed local units measuring and tracking the state of dynamic system.} in CPS. 
%From this viewpoint, CPS security, vulnerability, and resilience to failures/attacks is tied with the observability and discoverability of the sensor network  \cite{roy2012vulnerable,asilomar14,roy2012security}. 
 
\textit{Contributions:} The questions that we address here are as follows: how the failure of a sensor affects the \textit{distributed} observability in the single time-scale distributed estimation? And, what are the counter-measures to recover for this failure? These questions are to a great extent unexplored in the literature. It is known that if the sensor is essential for estimation its failure implies loss of necessary information and loss of system observability. We propose to overcome this problem by placing new sensors to measure new states in the dynamical system. The idea is to infer \textit{equivalent} information of the system to regain distributed observability and to stabilize the Mean Squared Error (MSE) of estimation.
The sensors are classified into $\alpha$-type and $\beta$-type based on their role in distributed estimation, see e.g., \cite{jstsp,jstsp14} for details. Using a graph-theoretic approach, we  show that if the failed sensor is $\alpha$-type measuring state $x_i$, the observability is recovered by adding a new sensor measuring a state $x_j$ sharing a \textit{contraction} with $x_i$ in system digraph. A polynomial order algorithm for contraction detection is provided. 
%The replaced $\alpha$-sensor directly shares its measurement of new state $x_j$ with all other sensors. 
On the other hand, if the failed sensor is $\beta$-type measuring state $x_i$, the estimation is recovered by adding a new sensor measuring state $x_j$ sharing a Strongly Connected Component (SCC) with $x_i$ in system digraph. 
%In this case, the replaced $\beta$-sensor shares its \textit{state prediction} with the neighboring sensors over a strongly-connected network. 
Note that, the proposed methods are graph-theoretic and irrespective of the numerical system parameters, that are particularly significant for large-scale system application.

%The paper outline is as follows. Section~\ref{sec_system} states  the system and observability notions. Particularly the graph theoretic approach towards estimation and observability is discussed. Section~\ref{sec_dist} presents the distributed estimator, its stability criteria, and the results on network connectivity. Section~\ref{sec_fail} provides the results on sensor failure recovery. Section~\ref{sec_sim} gives an academic example illustrating the results. Finally, Section~\ref{sec_con} concludes the paper.    
\section{Problem Formulation} \label{sec_prob}
Consider estimation of an LTI system with measurements:
\begin{eqnarray}\label{eq_sys1}
\mb{x}_{k+1} &=& A\mb{x}_k + \mb{v}_k,
\\\label{eq_sys2}
\mb{y}_k^j &=& H_j\mb{x}_k + \mb{r}_k^j,\qquad j\in \{1,...,m\}.
\end{eqnarray}
where~$\mb{x}=[x_{1}~\ldots~x_{n}]^\top\in\mbb{R}^n$ is state-space,~$\mb{y}=[y_1,\ldots,y_m]\in\mbb{R}^m$ as measurement vector, ~$\mb{v}$ and~$\mb{r}$, as noise variables with standard assumptions on Gaussianity and independence, and~$k$ is the discrete-time index. Based on Kalman \cite{kalman:61}, estimation of the noise-corrupted LTI model in Eqs.~\eqref{eq_sys1}-\eqref{eq_sys2} renders bounded estimation error if and only if the system is \textit{observable (detectable)}\footnote{In this paper, it is assumed that there are no stable unobservable modes in the system. Therefore, detectability and observability are equivalent.}. Observability ensures that the entire system state,~$\mb{x}$, can be uniquely recovered from the noise-corrupted measurements,~$\mb{y}$.

In this paper, single time-scale distributed estimation is addressed. In single time-scale distributed estimation, both system dynamics and distributed estimator evolve at the same time-scale \cite{das2016consensus,das2015distributed,sayedtu12,sayed11,kar2013consensus,jstsp,jstsp14, sauter:09,nuno-suff.ness}. This method is privileged with tracking faster system dynamics and imposing less communication burden. This is contrary to the two time-scale method where information-fusion (consensus) is performed faster than system dynamics \cite{olfati:05,Msechu:08}. In the single-time estimation method, two types of information fusion are performed: (i) prediction-fusion, i.e. sharing sensors' prediction of the state over a communication network; and (ii) measurement fusion, i.e. sharing sensors' measurements over a communication network. Either one of these two types of information-fusion (for example only prediction fusion \cite{nuno-suff.ness} or only measurement fusion  \cite{sauter:09}\footnote{Note that the case of only measurement fusion needs a fully-connected sensor network, and therefore, is a trivial case (not considered in this paper).}) or both of them \cite{jstsp,jstsp14} can be applied. Accordingly, the sensors are classified based on performing each type of information-fusion as follows:
\begin{defn}\label{def_sensor}
	Sensors are type-$\alpha$ if they are required to share their measurement over communication network. Sensors are type-$\beta$ if they are required to share their state-prediction over communication network.
\end{defn}
In this direction, the communication network of sensors  are defined as follows:
\begin{defn}\label{def_G}
	Sensors perform prediction-fusion over network $\mc{G}_\beta$, with $\mc{N}_\beta$ defined as the neighborhood of nodes and matrix $W$ as its adjacency matrix\footnote{Note that the weights in the adjacency matrix are defined such that $W$ is row-stochastic, see \cite{jstsp,jstsp14,nuno-suff.ness} for details.}. Sensors perform measurement fusion over network $\mc{G}_\alpha$, with $\mc{N}_\alpha$ defined as the neighborhood of nodes.
\end{defn}  
Further define $D_H$ as:
\begin{eqnarray}\nonumber
D_H = blockdiag \left( \sum_{j\in\mc{N}_\alpha (1)} H_j^TH_j \ldots \sum_{j\in\mc{N}_\alpha (m)} H_j^TH_j\right)
\end{eqnarray}
%\subsection{Error Stability}
%The main theorem on stability of the estimation error~\eqref{eq_err1} is given in the following:

\begin{theorem} \label{thm_stability}
	Given a system matrix $A$, measurement matrices $H_i, i\in\{1,...,m\}$, communication network $\mc{G}_\beta$ and $\mc{G}_\alpha$, if $(W\otimes A, D_H)$ is observable\footnote{We refer $(W\otimes A, D_H)$-observability as \textit{distributed observability}.} then there exist a  gain matrix $K_k$ such that the distributed estimation achieves asymptotic omniscience and the error dynamics achieves global asymptotic stability on Mean Squared Error (MSE).
\end{theorem}
%\begin{proof}
	 The detailed proof is given in \cite{jstsp,jstsp14,globalsip14}. The proof for the special case of (only) prediction fusion is provided in \cite{nuno-suff.ness,sayed11} and (only) measurement fusion in \cite{sauter:09}.
%\end{proof}

\begin{rem} \label{rem_alphabeta}
	The communication network $\mc{G}_{\alpha}$ is  defined as a hub network, where every $\alpha$-sensor directly shares its information with all other sensors. On the other hand, the communication network $\mc{G}_{\beta}$ is a strongly-connected network. For more details see \cite{jstsp,jstsp14,sauter:09,nuno-suff.ness,sayed11} 
\end{rem}

The problem we address in this paper is when one (or more) sensor fail to measure a state, and therefore, the system is not observable to the sensor network, i.e. the distributed estimator loses distributed observability. The countermeasure is to add a new sensor in the sensor network to recover for loss of observability, while the method for observability recovery and necessary connectivity in the sensor network differs depending on the type of sensor ($\alpha$  or $\beta$). Overall, the distributed observability recovery in sensor networks is defined in the following.

\textbf{Problem Statement:} Given system matrix $A$, measurement matrices $H_i, i\in\{1,...,m\}$, communication network $\mc{G}_\beta$ and $\mc{G}_\alpha$, if sensor taking measurement $H_j$ fails then how to recover $(W\otimes A, D_H)$-observability with a new sensor measurement $H_{j'}$.\footnote{Note that here without loss of generality one sensor failure is considered. The results can be easily extended to the case of more than one sensor failure.} 

\section{Graph-theoretic Observability}\label{sec_obsrv}
To solve the problem, instead of the algebraic Grammian observability test, an alternative graph-theoretic approach is applied over \emph{system digraph}. We first introduce some graph notions to present the main result on graph-theoretic observability and classification. Let~$\mc{X}=\{x_1,\ldots,x_n\}$ and~$\mc{Y}=\{y_1,\ldots,y_m\}$ denote the set of state and measurement nodes, respectively. The system digraph is a \emph{directed adjacency graph} defined as~$\mc{G}_A = (\mc{X} \cup \mc{Y},\mc{E})$, with~$\mc{E}$ as the set of edges/links;  
a link,~$x_j {\rightarrow} x_i$, in~$\mc{E}$ exists from~$x_j$ to~$x_i$ if~$a_{ij}\neq0$. Similarly, a  link,~$x_j {\rightarrow} y_i$, in~$\mc{E}$ exists from~$x_j$ to~$y_i$ if~$h_{ij}\neq0$. A \emph{path} from~$x_j$ to~$x_i$ (or~$y_i$) is denoted as~$x_j\overset{\scriptsize\mbox{path}}{\longrightarrow} x_i$. A path is called \emph{$\mc{Y}$-connected} (denoted by~$\overset{\scriptsize\mbox{path}}{\longrightarrow} \mc{Y}$) if it ends in $\mc{Y}$. A \emph{cycle} is a path starting and ending at the same node. A cycle-family is a set of cycles which are mutually disjoint and do not share any node. Similarly, a path and a cycle are disjoint if they do not share any node. More details on this construction can be found in~\cite{jstsp,jstsp14,godsil}. 

\begin{theorem}\label{woude_thm}
A system is structurally-observable if and only if in its digraph~$\mc{G}_A$:
\begin{enumerate}[(i)]

\item Every state~$x_i$ is included in a~$\mathcal{Y}$-connected path, i.e.~$x_i \overset{\scriptsize\mbox{path}}{\longrightarrow} \mc{Y}, \forall i \in \{1,\ldots,n\}$;

\item  Every state~$x_i$ is included in a disjoint family of~$\mathcal{Y}$-connected paths and cycles spanning $\mc{X}$.
\end{enumerate}

\end{theorem}
%\begin{proof}
	See the proof in \cite{liu-pnas}, and the proof for dual problem of structural controllability in~\cite{lin,Liu-nature}.
%\end{proof}
 
Next we review some graph notions on the bipartite representation of system to find the necessary states for observability.  A bipartite graph,~$\Gamma=(\mc{V}^+,\mc{V}^-,\mc{E}_\Gamma)$, consists of two disjoint sets of nodes~$\mc{V}^+$ and~$\mc{V}^-$ with set of links ~$\mc{E}_\Gamma$ starting in~$\mc{V}^+$ and ending in~$\mc{V}^-$. The set of links and nodes are defined as follows: $\mc{V}^+=\mc{X}$,~$\mc{V}^- = \mc{X}$, and the link set,~$\mc{E}_{\Gamma_A},$ defined as the collection of~$(v_j^-,v_i^+)$, if ~$(v_j,v_i) \in \mc{E}_A$. Define a matching, denoted by~$\underline{\mc{M}}$, in the bipartite graph,~$\Gamma_A$, as the subset of links chosen such that they do not share an end-node in~$\mc{V}^-$ or a start-node in~$\mc{V}^+$ . This implies that the links in~$\underline{\mc{M}}$ are mutually disjoint. The size (cardinality) of the matching, denoted by~$|\underline{\mc{M}}|$, is the number of links in~$\underline{\mc{M}}$.  A \textit{maximum matching},~$\mc{M}$, is a matching with maximum size.  
Note that maximum matching is \textit{not unique}, in general. Given a  matching,~$\underline{\mc{M}}$, denote by~$\partial \underline{\mc{M}}^+$ the set of matched nodes incident to~$\underline{\mc{M}}$ in~$\mc{V}^+$. The set of \textit{unmatched nodes}, denoted by~$\delta \underline{\mc{M}}$, is the set of nodes in~$\mc{V}^+$ not belonging to~$\partial \underline{\mc{M}}^+$, i.e.,~$\delta \underline{\mc{M}} = \mc{V}^+ \backslash \partial \underline{\mc{M}}^+~$.
Next, define a Strongly Connected Component (SCC) as a subgraph in which every node is reachable via a directed path to every other node, i.e. for every $v_i,v_j \in SCC$ we have $v_i\overset{\scriptsize\mbox{path}}{\longrightarrow} v_j$. A parent SCC
is defined as a SCC with no outgoing links to any nodes not belonging to that SCC, i.e. for every $v_i \in SCC^p$ there is no $v_j \notin SCC^p$ such that $v_i{\longrightarrow} v_j$.
Having these graph-theoretic notions we are ready to present the following theorem: 
\begin{theorem} \label{thm_alphabeta}
	The system \eqref{eq_sys1} is observable if and only if the set of following state nodes in system digraph are measured:
	
	(i) every unmatched state node $x_j \in \delta \mc{M}$, 
	
	(ii) at least one state in every parent SCC in $\mc{G}_A$.
	
\end{theorem}
%\begin{proof}
	The proof is given in our previous work \cite{jstsp,jstsp14}. The proof of part (i) for SC networks (for dual case of controllability) is given in \cite{Liu-nature}.
%\end{proof}
%using this theorem we classify the sensors accordingly:

Following Definition~\ref{def_sensor} the sensors are classified according to the following lemma: 
\begin{lem} \label{lem_alphabeta}
	A sensor measuring an unmatched state node $x_j \in \delta \mc{M}$ in system digraph is  $\alpha$-type. A sensor measuring a state in a parent SCC is  $\beta$-type.
\end{lem}
The proof is given in our previous work \cite{jstsp,jstsp14}. 

\section{Sensor Failure Recovery} \label{sec_fail}
Sensor failures are prevalent in sensor networks due to harsh environmental conditions, possible adversary attacks, and the nature of sensing devices \cite{yick2008wireless}. In this section, we investigate the countermeasures to recover distributed observability ($(W \otimes A, D_H)$-observability) under sensor failure. We separately analyze sensor failure in $\mc{G}_{\alpha}$ for measurement update and in $\mc{G}_{\beta}$ for prediction fusion and the countermeasures to recover for distributed observability by adding equivalent sensor measurements.   

\subsection{Type-$\alpha$ Sensor Failure}
Recall from Lemma~\ref{lem_alphabeta} that an $\alpha$-sensor measures an unmatched state in system digraph and share its measurement (as a hub) over $\mc{G}_{\alpha}$. Having a failed $\alpha$-sensor implies the loss of (distributed) observability of an unmatched state. Let $H_{\alpha}$ represents the failed measurement, therefore, according to definition, $D_H$ must be recovered. The solution is by adding a new state measurement $H_{\alpha'}$ of \textit{structurally equivalent} unmatched state. The equivalent unmatched state nodes are defined by the set of \textit{contractions} in system digraph. 

\textbf{Contraction sets:}
Denote by~$\Gamma^{\underline{\mc{M}}}_A=(\mc{V}^+,\mc{V}^-,\mc{E}_{\Gamma_A})$ the auxiliary graph associated to a matching,~$\underline{\mc{M}}$, and the bipartite graph,~$\Gamma_A$. Given~$\Gamma_A$, this graph is made by reversing all the links in~$\underline{\mc{M}}$ while preserving the direction of links in~$\mc{E}_{\Gamma_A} \backslash \underline{\mc{M}}$. Given the matching~$\underline{\mc{M}}$ and auxiliary graph~$\Gamma^{\underline{\mc{M}}} _A$, an ${\underline{\mc{M}}}$-alternating path,~$\mc{Q}_{\underline{\mc{M}}}$, is constructed by sequence of links alternating between unmatched links,~$\mc{E}_{\Gamma_A} \backslash \underline{\mc{M}}$, and matched links,~$\underline{\mc{M}}$. The sequence starts with an unmatched link in~$\mc{E}_{\Gamma_A} \backslash \underline{\mc{M}}$ from a free node in~$\delta \underline{\mc{M}}$ and every second link in~$\underline{\mc{M}}$. Given the matching~$\underline{\mc{M}}$ and auxiliary graph~$\Gamma^{\underline{\mc{M}}} _A$, an ${\underline{\mc{M}}}$-augmenting path,~$\mc{P}_{\underline{\mc{M}}}$, is an alternating path with start node and end node in~$\delta \underline{\mc{M}}$. Then, the contraction in the system is defined as follows:

\begin{defn} 
	Given a maximum matching, $\mc{M}$, for every unmatched node,~$v_j \in \delta \mc{M}$, in auxiliary graph,~$\Gamma^\mc{M} _A$, find the  set of all nodes in $\mc{V}^+$ reachable by alternating paths,~$\mc{Q}_{\mc{M}}$, from~$v_j$. This is called the set of contraction nodes, denoted by $\mc{C}$. 
\end{defn}
\begin{algorithm} \label{alg_cont}
	\textbf{Given:} System digraph $\mc{G}_A$ 
	
	\KwResult{Contractions $\{\mc{C}_1,...,\mc{C}_l\}$}
%	\textbf{Initialization:} \; 
Construct $\Gamma_A$\;
Find a matching $\underline{\mc{M}}$ \;
Construct $\Gamma^{\underline{\mc{M}}}_A$ \;
\While{augmenting path $\mc{P}_{\underline{\mc{M}}}$ exist}{
	\For{nodes in $\delta \underline{\mc{M}}$}{  
		Find $\mc{P}_{\underline{\mc{M}}}$ \;
		$\underline{\mc{M}} = \underline{\mc{M}} \oplus \mc{P}_{\underline{\mc{M}}}$ \;
	}
}
	Construct $\Gamma^{\mc{M}}_A$ \;
	\For{nodes in $\delta \mc{M}$}{  
		Find alternating paths $\mc{Q}_{\mc{M}}$ in $\Gamma^{\mc{M}}_A$ \;
		Put all nodes in $\mc{V}^+$ reachable by $\mc{Q}_{\mc{M}}$ in $\mc{C}_i$\;}
	
	\textbf{Return} $\mc{C}_i, i = \{1,...,l\}$\;\
	
	\caption{Contraction Detection Algorithm} %: modified version of Hopcroft-Karp Algorithm in \cite{murota,hopcraft}}
\end{algorithm}
Notice that in Algorithm~\ref{alg_cont}, $\oplus$ represents the XOR operator in set theory. As a result of this operator, each augmenting path increases the size of the matching by one till it reaches the maximum matching. 
More detailed explanations and definitions on contractions and related topics is available in \cite{murota}.

\begin{rem} 
	Every state node in a contraction $\mc{C}_i$ represents an unmatched node for a choice of maximum matching, $\mc{M}$ \cite{murota,hopcraft,jstsp14}. Further, the size of contraction determines number of possible options of equivalent states for sensor recovery.  
\end{rem}

\begin{rem}
	Note that the connectivity of the new sensor $\alpha'$ in $\mc{G}_{\alpha}$ is similar to the connectivity of the failed sensor $\alpha$.
\end{rem}

\subsection{Type-$\beta$ Sensor Failure}
Recall from Lemma~\ref{lem_alphabeta} that a $\beta$-sensor  measures a state in a parent SCC and shares its state prediction on a strongly-connected graph $\mc{G}_{\beta}$. Having a failed $\beta$-sensor implies the loss of necessary information on state prediction. Let $H_{\beta}$ represents the failed sensor measurement, therefore according to Definition~\ref{def_G}, the $W \otimes A$ must be recovered. This is by an \textit{equivalent} state measurement $H_{\beta'}$ providing equivalent state prediction. These equivalent states are defined by the set of states sharing a SCC  as discussed in Section~\ref{sec_obsrv}.  Theorem~\ref{thm_alphabeta} part-(ii) states that measurement of at least one state (any state) from each parent SCC is necessary for observability. Therefore, losing the measurement of one state in a parent SCC, the observability is recovered by measuring another state in the same SCC. This is done via well-known graph-theoretic algorithms for SCC classification, namely Depth First Search (DFS) \cite{algorithm} or Tarjan algorithm \cite{tarjan}. These algorithms are widely provided in related literature \cite{algorithm,tarjan,murota} and are omitted here due to space limitation.

\begin{rem}
	The size of parent SCC determines the number of equivalent states. If the parent SCC is in the form of a \textit{self-cycle}, it only includes one state node and therefore there is no equivalent state measurement to recover for estimation. 
\end{rem}
\begin{rem}
Note that the new sensor $\beta'$ follows similar connectivity as the failed sensor $\beta$, i.e. it shares its prediction over a strongly-connected graph $\mc{G}_{\beta}$.
\end{rem}
\section{Illustrative Example and Simulation} \label{sec_sim}
\begin{figure}[t]
	\centering
	\includegraphics[width=1.45in]{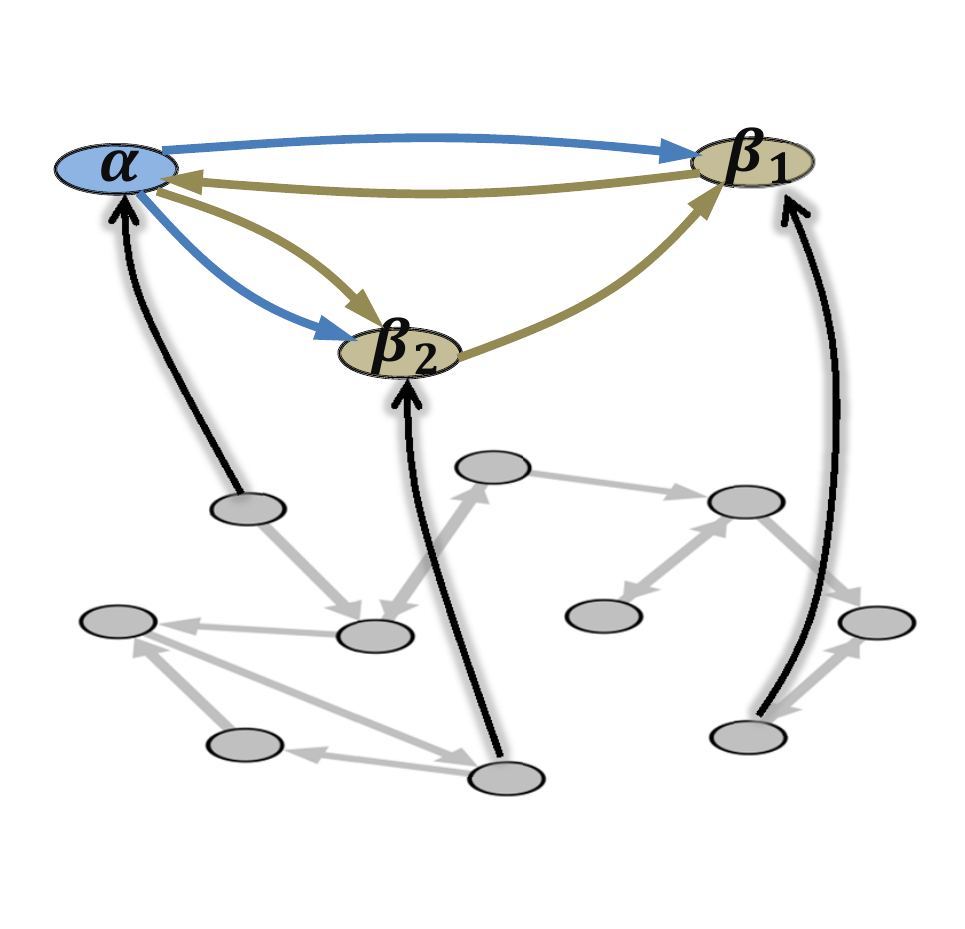}
	\centering
	\includegraphics[width=1.7in] {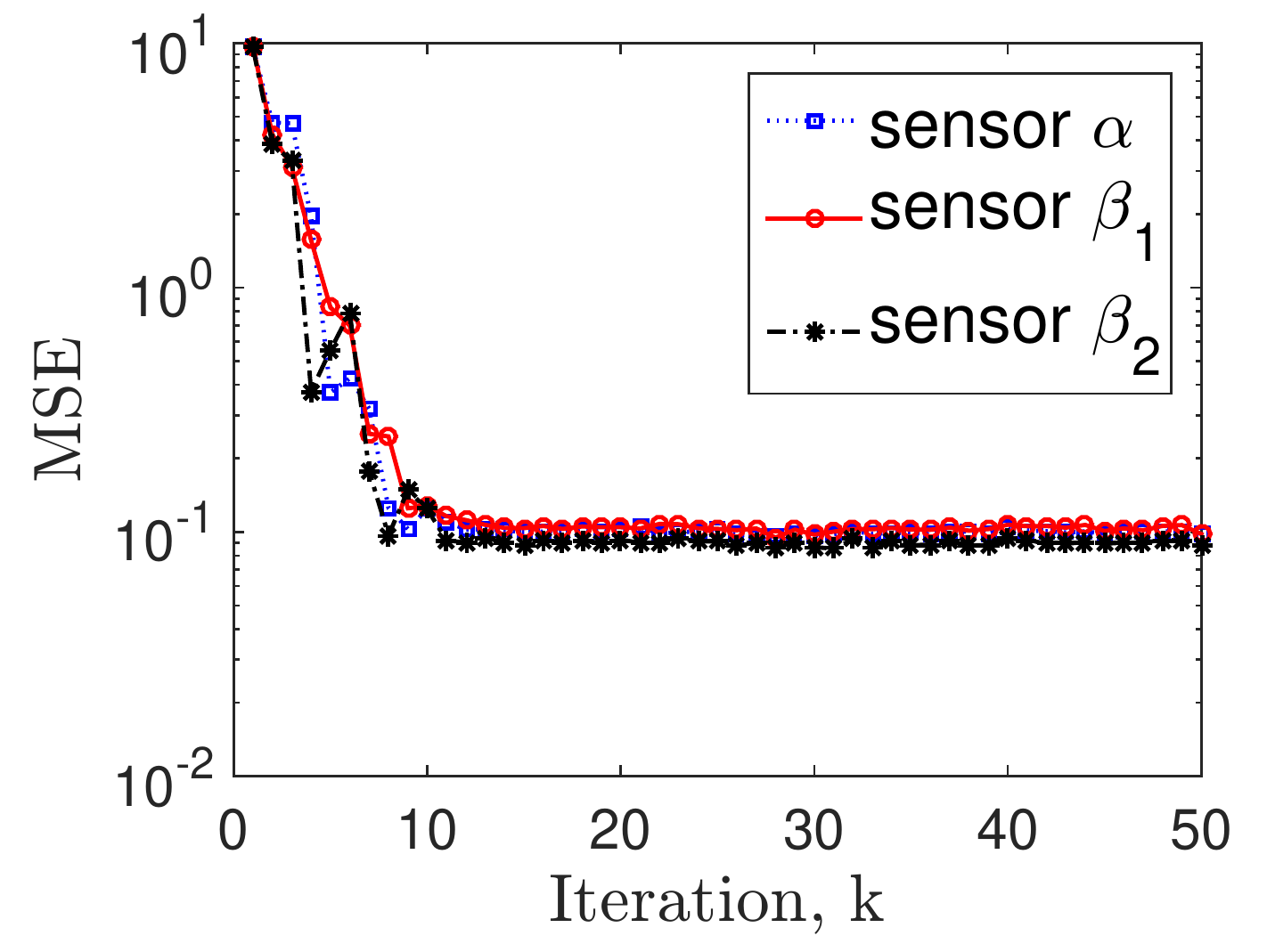}
	\caption{(Left) The system digraph (the graph at bottom) is monitored by a sensor network (the graph at top) which includes $\mc{G}_{\alpha}$ in blue and $\mc{G}_{\beta}$ in green. The black arrows from system digraph to the sensor network represent measurement of states by sensors. (Right) The time-evolution of the MSE at sensors using the distributed estimator \eqref{eq_pf}-\eqref{eq_mf}.}
	\label{fig_sim}
\end{figure}

We consider a discrete-time dynamic system of $n=10$ states with random entries in $A$, such that the given dynamic system is unstable with spectral radius $\rho(A)=1.1>1$. This ensures that the stability of tracking error  is not trivial. Following the results of Theorem~\ref{thm_alphabeta}, we find the SCCs and unmatched node of the system. From Theorem~\ref{thm_alphabeta} number of necessary sensors for observability is $m=3$, one measuring the unmatched state and the other two each measuring a state in a parent SCC (see Fig.\ref{fig_sim}-Left). Zero-mean Gaussian noise with covariance $0.25^2$ for both measurement and system is considered. Following distributed estimator is applied:
\begin{eqnarray}\label{eq_pf}
\widehat{\mb{x}}^i_{k|k-1} &=& \sum_{j\in \mathcal{N}_\beta(i)} w_{ij}A\widehat{\mb{x}}^j_{k-1|k-1}, \\ \label{eq_mf}
\widehat{\mb{x}}^i_{k|k} &=&\widehat{\mb{x}}^i_{k|k-1} + K_k^i \sum_{j\in \mc{N}_\alpha(i)}H_j^\top \left(\mb{y}^j_k-H_j\widehat{\mb{x}}^i_{k|k-1}\right).
\end{eqnarray}
with $\widehat{\mb{x}}^i_{k|m}$ defined as estimation of $\mb{x}$ by sensor $i$ at step $k$ given the measurements up to time $m$. The first step~\eqref{eq_pf} is \textit{prediction fusion} and the latter step \eqref{eq_mf} is \textit{measurement fusion}. 
For this estimator the error dynamics is given as the following equation \cite{jstsp}:
\begin{eqnarray}\label{eq_err1}
\mb{e}_{k} = (W\otimes A - {K}_kD_H(W\otimes A))\mb{e}_{k-1} +
\mb{q}_k,
\end{eqnarray}
where $\mb{e}_{k} = [(\mb{e}_{k}^1)^\top \ldots (\mb{e}_{k}^m)^\top]^\top$ is global error, and $\mb{q}_k$ collects the noise terms.  Note that, to have a fully distributed estimator, the gain matrix $K_k$ is block diagonal where the $i$th diagonal block~$K_k^i$ is the gain matrix for the estimator $i$.

For distributed estimation and stability of the MSE according to Remark~\ref{rem_alphabeta}, $\alpha$-sensor directly shares its measurement of system unmatched state and two $\beta$-sensors share their state predictions over a strongly-connected network. This network setting guarantees that the pair $(W\otimes A, D_H)$ is observable, for which a block-diagonal gain matrix $K$ may be found using the procedure described in \cite{rami:97,pang:95}, also see our prior work in \cite{jstsp}. 
The weights of the prediction fusion network are chosen randomly, but such that $W$ remains a stochastic matrix. The simulation is performed over $1000$ Monte-Carlo trials and the time-evolution of the Mean Squared Error (MSE) at each sensor is shown in Fig.\ref{fig_sim}-Right.

\begin{figure}[]
	\centering
	\includegraphics[width=1.45in]{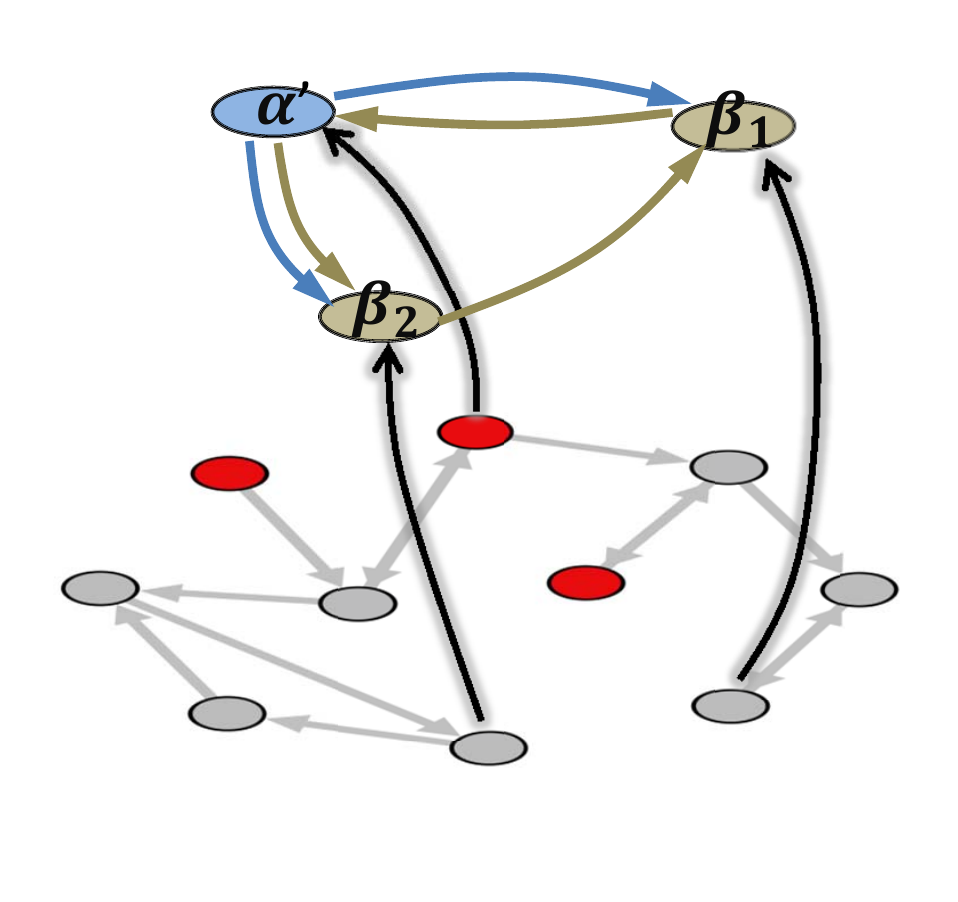}
	\includegraphics[width=1.7in] {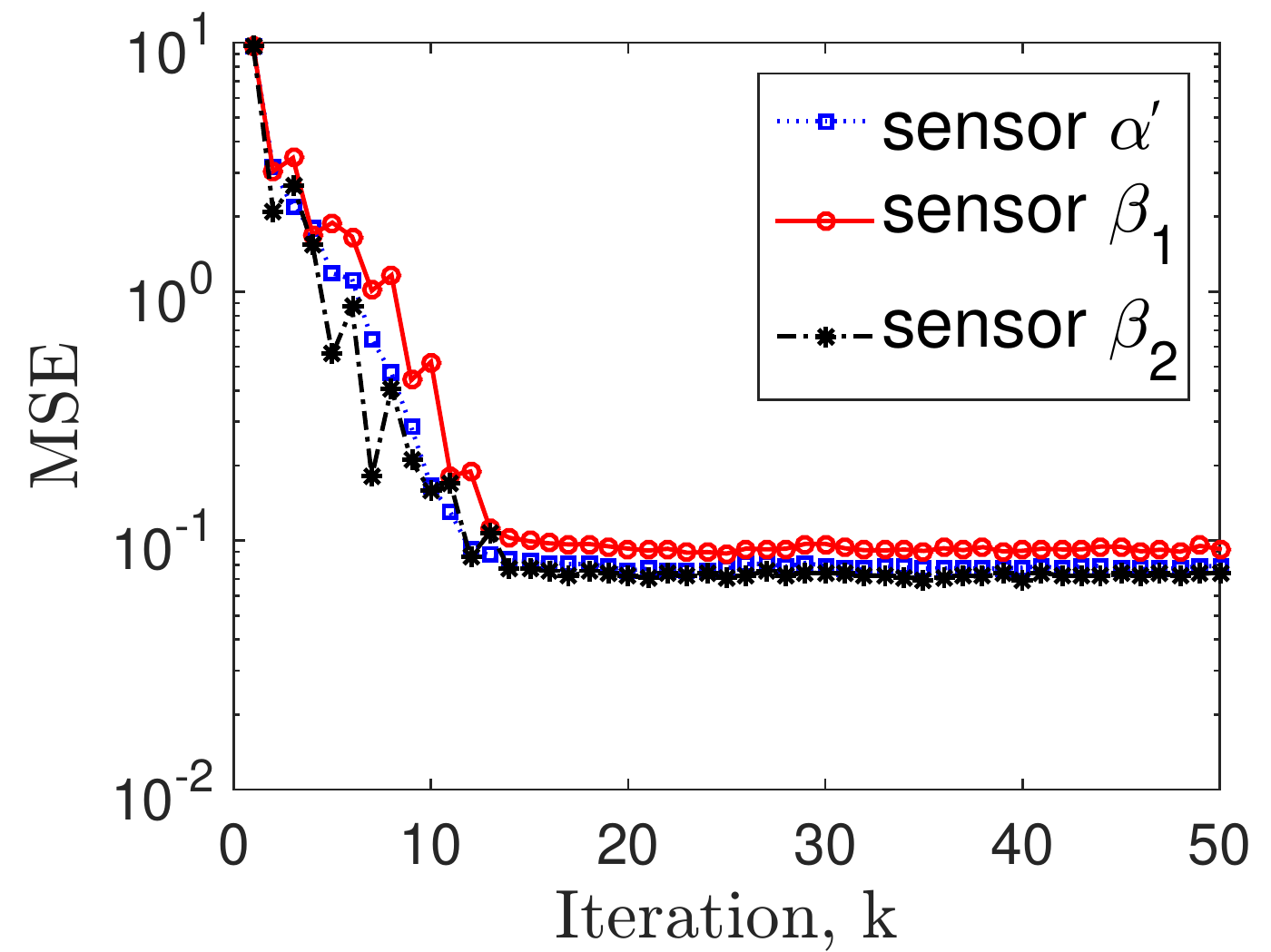}
	\includegraphics[width=1.44in]{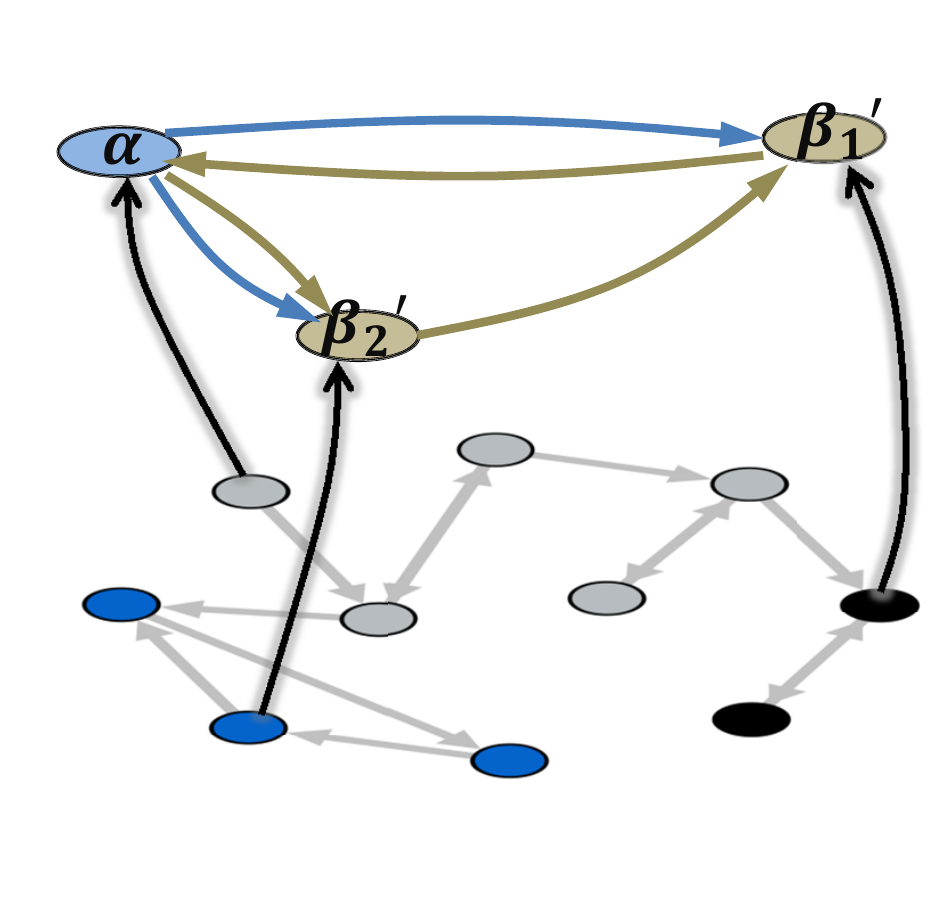}
	\includegraphics[width=1.7in] {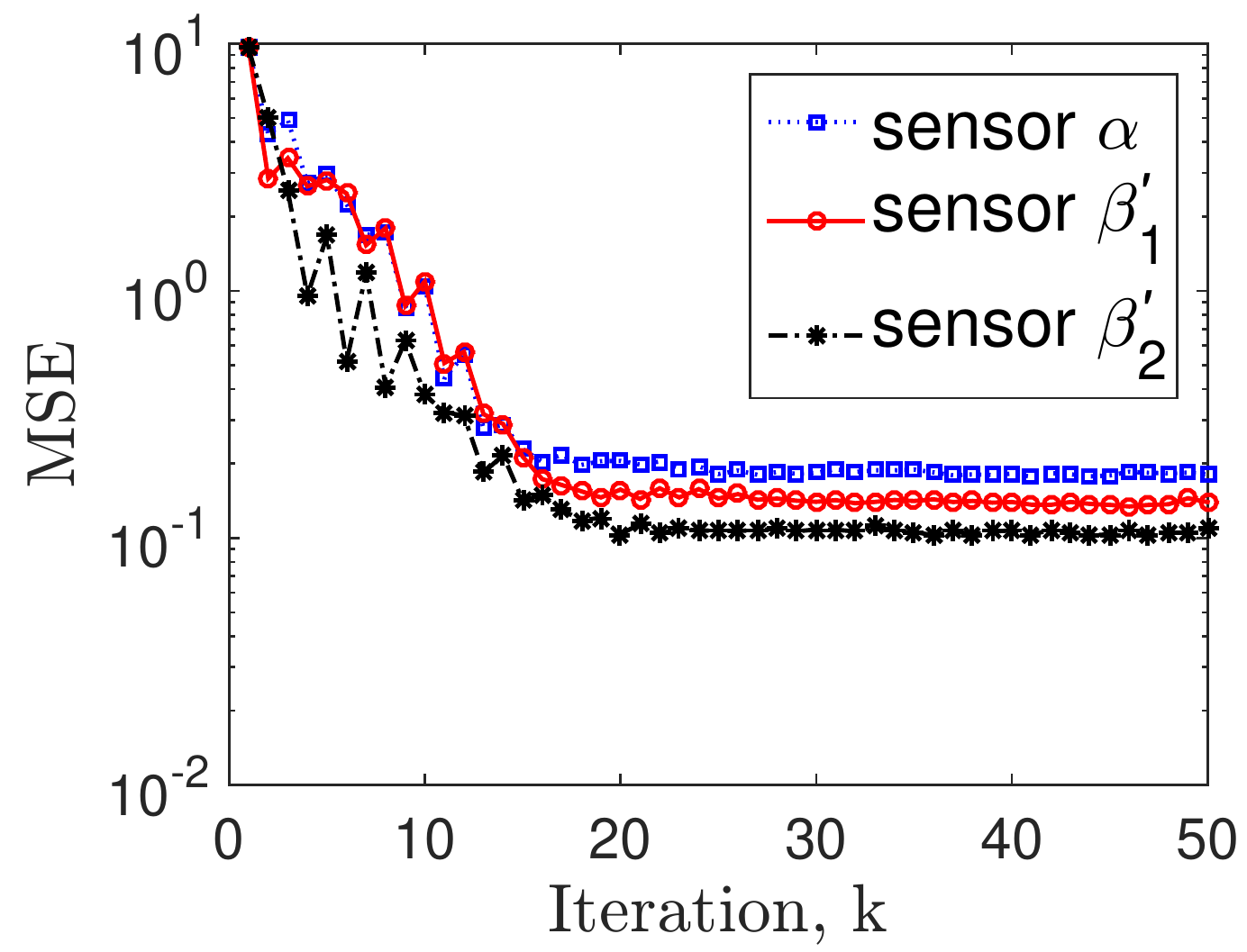}
	\caption{(TopLeft) Recovered sensor network by adding new $\alpha$-sensor measuring an equivalent state from the set of contraction nodes (shown in red). (TopRight) The MSE evolution in time for three sensors. (BottomLeft) Recovered sensor network by adding new $\beta$-sensors measuring equivalent states sharing SCC. (BottomRight) The time-evolution of MSE for all sensors.  }
	\label{fig_sim1}
\end{figure}

Next, we assume that the $\alpha$-sensor fails; therefore  $(W\otimes A, D_H)$ is not observable anymore and the tracking MSE goes unbounded for all sensors. As a countermeasure to recover for this, another sensor $\alpha'$ measuring an \textit{equivalent} unmatched state node  needs to be added. In Fig.\ref{fig_sim1}-TopLeft, using Algorithm~\ref{alg_cont} the set of equivalent unmatched nodes (contraction nodes) are shown in red color. Placing a sensor on the equivalent state and sharing its measurement directly with two other sensors recovers the distributed observability. This implies the stability of MSE for all sensors as shown in Fig.\ref{fig_sim1}-TopRight. Note that the design procedure for gain matrix $K$ is the same, as in \cite{jstsp,rami:97,pang:95}. Next, consider both $\beta$-sensors are failed. Therefore, sensor measurements from \textit{equivalent} states (sharing SCC) are required. This is given in Fig.\ref{fig_sim1}-BottomLeft, as states sharing a parent SCC are shown in different colors using DFS algorithm. Measuring an equivalent state in each SCC and sharing state predictions over a  strongly-connected network guarantee the stability of  MSE as shown in Fig.\ref{fig_sim1}-BottomRight. 

%Notice that any strongly-connected network works as $\mc{G}_\beta$. For example, reversing the green links in the sensor network of Fig.\ref{fig_sim} and \ref{fig_sim1} also gives a strongly-connected  network and ensures MSE stability.

\section{Concluding Remarks} \label{sec_con}
Distributed estimation recovery via equivalent sensor measurements is addressed. Graph-theoretic algorithms are proposed to find the equivalent states for recovery of two types of sensors.
The complexity of the Contraction Detection algorithm for $\alpha$-sensor recovery is $\mc{O}(n^{2.5})$. For SCC classification and $\beta$-sensor recovery the  complexity is $\mc{O}(n^2)$. The polynomial order solution and the graph theoretic approach supports application  in large-scale  systems. 

\bibliographystyle{IEEEbib}
\bibliography{bibliography}
\end{document}